\documentclass[sigconf]{acmart}

\usepackage{booktabs} 
\usepackage{alltt} 
\usepackage{geometry} 
\usepackage{tikz} 
\usepackage{balance}
\usepackage{xcolor}
\newcommand{\rev}[1]{\textcolor{black}{#1}}

\setcopyright{rightsretained}

\acmDOI{10.1145/3299768}

\acmConference[Communications of the ACM ]{CACM}{August 2019}{Vol. 62 No. 8, Pages 82-91}
\acmYear{2019}
\copyrightyear{2019}

\acmPrice{15.00}

\acmBooktitle{Communications of the ACM, August 2019, Vol. 62 No. 8, Pages 82-91}

\begin{document}
\title{\rev{The History of Digital Spam}}

\author{Emilio Ferrara}
\orcid{0002-1942-2831}
\affiliation{%
  \institution{University of Southern California\\Information Sciences Institute}
  \streetaddress{4676 Admiralty Way, \#1001}
  \city{Marina Del Rey}
  \state{CA}
  \postcode{90292}
}
\email{emiliofe@usc.edu}

\renewcommand{\shortauthors}{E. Ferrara}


%




\newcommand{\mybox}[4]{
    \begin{figure}[h!]
        \centering
    \begin{tikzpicture}
        \node[anchor=text,text width=.87\columnwidth, draw, rounded corners, line width=1pt, fill=#3, inner sep=5mm] (big) {\\\small#4};
%
        \node[draw, rounded corners, line width=.5pt, fill=#2, anchor=west, xshift=5mm] (small) at (big.north west) {#1};
    \end{tikzpicture}
    \end{figure}
}

\maketitle

\begin{figure*}
\includegraphics[clip = true, trim = 75 50 75 75, width=2.1\columnwidth]{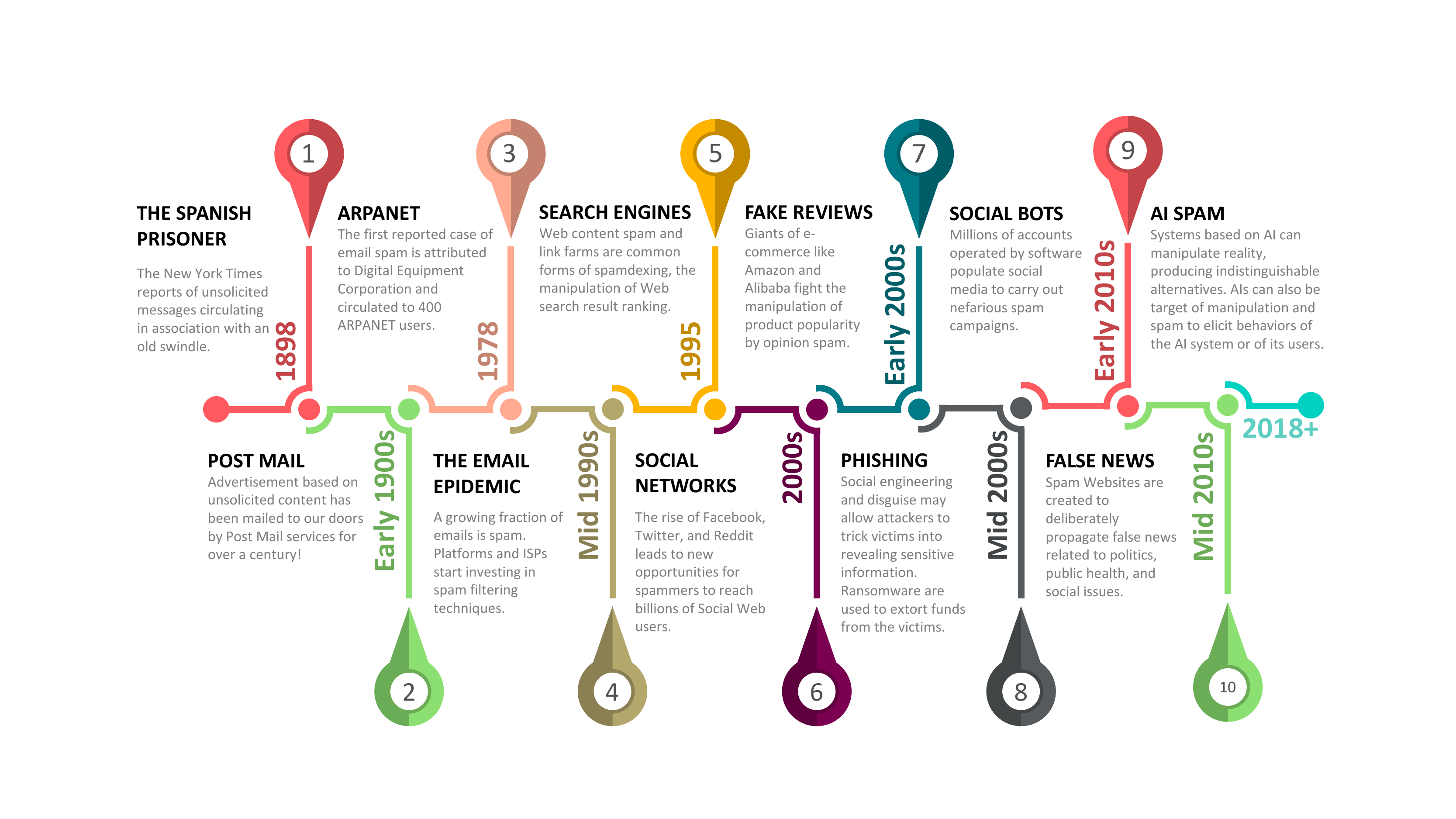}
\label{fig:spam}
\caption{Timeline of the major milestones in the history of spam, from its inception to modern days.}
\end{figure*}

\textit{Spam!}: \rev{that's what Lorrie Faith Cranor and Brian LaMacchia exclaimed in the title of a popular call-to-action article that} appeared twenty years ago on \textit{Communications of the ACM}~\cite{cranor1998spam}. And yet, despite the tremendous efforts of the research community over the last two decades to mitigate this problem, the sense of urgency remains unchanged, as emerging technologies have brought new dangerous forms of digital spam under the spotlight. Furthermore, when spam is carried out with the intent to deceive or influence at scale, it can alter the very fabric of society and our behavior. In this article, I will briefly review the history of digital spam: starting from its quintessential incarnation, spam emails, to modern-days forms of spam affecting the Web and social media, the survey will close by depicting future risks associated with spam and abuse of new technologies, including Artificial Intelligence (e.g., Digital Humans). After providing a taxonomy of spam, and its most popular applications emerged throughout the last two decades,  I will review technological and regulatory approaches proposed in the literature, and suggest some possible solutions to tackle this ubiquitous digital epidemic moving forward.

\section{Types of Spam}

An omni-comprehensive, universally-acknowledged  definition of digital spam is hard to formalize. \rev{Laws and regulation attempted to define particular forms of spam, e.g., email (cf., 2003's \textit{Controlling the Assault of Non-Solicited Pornography and Marketing Act}.)}
However, nowadays, spam occurs in a variety of forms, and across different techno-social systems. Each domain may warrant a slight different definition that suits what spam is in that precise context: \rev{some features of spam in a domain, e.g., \textit{volume} in mass spam campaigns, may not apply to others, e.g., carefully targeted phishing operations.}

In an attempt to propose a general taxonomy, I here \textit{define digital spam as the attempt to abuse of, or manipulate, a techno-social system by producing and injecting unsolicited, and/or undesired content aimed at steering the behavior of humans or the system itself, at the direct or indirect, immediate or long-term advantage of the spammer(s)}. 

This broad definition will allow me to track, in an inclusive manner, the evolution of digital spam across its most popular applications, starting from spam emails to modern-days spam. For each highlighted application domain, I will dive deep to understand the nuances of different digital spam strategies, including their intents and catalysts and, from a technical standpoint, how they are carried out  and how they can be detected.

Wikipedia provides an extensive list of domains of application: \begin{alltt}\small
``While the most widely recognized form of spam is email spam, 
the term is applied to similar abuses in other media: instant 
messaging spam, Usenet newsgroup spam, Web search engine spam,
spam in blogs, wiki spam, online classified ads spam, mobile 
phone messaging spam, Internet forum spam, junk fax
transmissions, social spam, spam mobile apps, television 
advertising and file sharing spam''. 
\end{alltt}

\begin{flushright}
(cf. \url{https://en.wikipedia.org/wiki/Spamming})
\end{flushright}

\begin{table}[b]
\begin{tabular}{@{}lll@{}cc@{}}
Spam Type & Start & Today's Volume & ML  & Ref\\
\hline
Email & 1978 & Billions x day & \checkmark & \cite{cranor1998spam}\\
Instant Messaging & 1997 & Millions x day & \checkmark & \cite{goodman2007spam} \\
Search Engine & 1998 & Unknown & \checkmark & \cite{spirin2012survey}\\
Wiki & 2001 & Thousands x day & - & \cite{adler2010detecting} \\
Opinion \& Reviews & 2005 & Millions across platforms & \checkmark & \cite{crawford2015survey} \\
Mobile Messaging & 2007 & Millions x day & \checkmark & \cite{almeida2011contributions} \\
Social Bots & 2010 & Millions across platforms & \checkmark & \cite{ferrara2016rise} \\
False News & 2016 & Thousands across Websites & - & \cite{vosoughi2018spread} \\
Multimedia & 2018 & Unknown & - & \cite{kim2018deep} \\
\end{tabular}
\caption{Examples of types of spam and relative statistics.}
\label{tab1}
\end{table}

Table \ref{tab1} summarizes a few examples of types of spam and relative context, including whereas there exist machine-learning solutions (ML) to each problem. Email is known to be historically the first example of digital spam (cf. Figure 1) and remains uncontested in scale and pervasiveness with billions of spam emails generated every day \cite{cranor1998spam}. In the late 1990s, spam landed on instant messaging (IM) platforms (SPIM) starting from AIM (AOL Instant Messenger\texttrademark) and evolving through modern-days IM systems such as WhatsApp\texttrademark, Facebook Messenger\texttrademark, WeChat\texttrademark, etc. A widespread form of spam that emerged in the same period was Web search engine manipulation: content spam and link farms allowed spammers to boost the position of a target Website in the search result rankings of popular search engines, by gaming algorithms like PageRank and the like. With the success of the Social Web \cite{hendler2008web}, in the early 2000s we witnessed the rise of many new forms of spam, including Wiki spam (injecting spam links into Wikipedia pages \cite{adler2010detecting}), opinion and review spam (promoting or smearing products  by generating fake online reviews \cite{liu2012sentiment}), and mobile messaging spam (SMS and text messages sent directly to mobile devices \cite{almeida2011contributions}). Ultimately, in the last decade, with the increasing pervasiveness of online social networks and the significant advancements in Artificial Intelligence (AI), new forms of spam involve social bots (accounts operated by software to interact at scale with Social Web users \cite{ferrara2016rise}), false news Websites (to deliberately spread disinformation \cite{vosoughi2018spread}), and multimedia spam based on AI  \cite{kim2018deep}.

In the following, I will focus on three of these domains: email spam, Web spam (specifically, opinion spam and fake reviews), and social spam (with a focus on social bots). Furthermore, I will highlight the existence of a new form of spam that I will call \textit{AI spam}. I will provide examples of spam in this new domain, and lay out the risks associated with it and possible mitigation strategies. 

\section{Flooded by junk emails}
\subsection{The Origins of Email Spam}
\citet{cranor1998spam}, in their 1998's \textit{Communications of the ACM} article, characterized the problem of \textit{junk emails}, or \textit{email spam}, as one of the earliest forms of digital spam.

Email spam has mainly two purposes, namely advertising (e.g., promoting products, services, or contents), and frauds (e.g.,  attempting to perpetrate scams, or \textit{phishing}). 
Neither ideas were particularly new or unique to the digital realm: advertisement based on unsolicited content delivered by traditional post mail (and, later, phone calls, including more recently the so-called ``robo-calls'') has been around for nearly a century. As for scams, the first reports of the popular \textit{advance-fee scam} (in modern days known as \textit{419 scam}, a.k.a. the \textit{Nigerian Prince scam}), called the \textit{Spanish Prisoner scam} were circulating in the late 1800s.\footnote{See \textit{The New York Times}, March 20, 1898: \url{https://www.nytimes.com/1898/03/20/archives/an-old-swindle-revived-the-spanish-prisoner-and-buried-treasure.html}}

\rev{The first reported case of digital spam occurred in 1978 and was attributed to \textit{Digital Equipment Corporation}, who announced their new computer system to over 400 subscribers of ARPANET, the precursor network of modern Internet (see Figure 1). The first mass email campaign occurred in 1994, known as the \textit{USENET green card lottery spam}: the law firm of \textit{Canter \& Siegel} advertised their immigration-related legal services simultaneously to over six thousand USENET newsgroups. This event contributed to popularizing the term \textit{spam}. Both the ARPANET and USENET cases brought serious consequences to their perpetrators as they were seen as egregious violations of common code of conduct in the early days of the Internet (for example, \textit{Canter \& Siegel} ran out of business and Canter was disbarred by the Arizona bar association.) However, things were bound to change as the Internet became an increasingly more pervasive technology in our society.}

\subsection{Email Spam: Risks and Challenges} 

The use of the Internet for distributing unsolicited messages provides unparalleled scalability, and unprecedented reach, at a cost that is infinitesimal compared to what it would take to accomplish the same results via traditional means \cite{cranor1998spam}. These three conditions created the ideal conjecture of economical incentives that made email spam so pervasive. 

In contrast to old-school post mail spam, digital email spam introduced a number of unique challenges \cite{cranor1998spam}: \textit{(i)} if left unfiltered, spam emails can easily outnumber legitimate ones, overwhelming the recipients and thus rendering the email experience from unpleasant to unusable; \textit{(ii)} email spam often contains explicit content that can hurt the sensibility of the recipients---depending upon the sender/recipient country's laws, perpetrating this form of spam could constitute a criminal offense;\footnote{E.g., see the U.S. Federal Law on Obscenity \url{https://www.justice.gov/criminal-ceos/citizens-guide-us-federal-law-obscenity}} \textit{(iii)} by embedding HTML or Javascript code into spam emails, the spammers can emulate the \textit{look and feel} of legitimate emails, tricking the recipients and eliciting unsuspecting behaviors, thus enacting scams or enabling phishing attacks~\cite{jagatic2007social}; finally, \textit{(iv)} mass spam operations pose a burden on Internet Service Providers (ISPs), which have to process and route unnecessary, and often large, amounts of digital junk information to millions of recipients---for the larger spam campaigns, even more. 

\rev{The Internet was originally designed by and for tech-savvy users: spammers quickly developed ways to take advantage of the unsophisticated ones. \textit{Phishing} is the practice of using deception and social engineering strategies by which attackers manage to trick victims by disguising themselves as a trusted entity \cite{jagatic2007social,chhabra2011phi}. The end goal of phishing attacks is duping the victims into revealing sensitive information for identity theft, or extorting funds via ransomware or credit card frauds. Email has been by far and large the most common vector of phishing attacks. In 2006, Indiana University carried out a study to quantify the effectiveness of phishing emails \cite{jagatic2007social}. The researchers demonstrated that a malicious attacker impersonating the university would have a 16\% success rate in obtaining the users' credentials when the phishing email came from an unknown sender; however, success rate arose to 72\% when the email came from an attacker impersonating a friend of the victim. }

\subsection{Fighting Email Spam}

Over the course of the last two decades, solutions to the problem of email spam revolved around implementing new regulatory policies, increasingly sophisticated technical hurdles,  and combinations of the two \cite{cranor1998spam}. Regarding the former, in the context of the U.S. or the European Union (EU), policies that regulate access to personal information (including email addresses), such as the EU's General Data Protection Regulation (GDPR) enacted in 2018, hinder the ability of bulk mailers based in EU countries to effectively carry out mass email spam operations without risks and possibly serious consequences. However, it has become increasingly more obvious that solutions based exclusively on regulatory affairs are ineffective: spam operations can move to countries with less restrictive Internet regulations. However, regulatory approaches in conjunction with technical solutions have brought significant progress in the fight against email spam.

From a technical standpoint, two decades of research advancements led to sophisticated techniques that strongly mitigate the amount of spam emails ending up in the intended recipients' inboxes. A number of review papers have been published that surveyed data mining and machine learning approaches to detect and filter out email spam \cite{caruana2012survey}, some with a specific focus on scams and phishing spam \cite{gupta2017fighting}.

In the \textit{Sidebar: Detecting Spam Emails}, I summarize some of the technical milestones accomplished in the quest to identify spam emails.
Unfortunately, I suspect that much of the state-of-the-art research on spam detection lies behind close curtains, mainly for three reasons: \textit{(i)} large email-related service providers, such as Google (Gmail\texttrademark), Microsoft (Outlook\texttrademark~, Hotmail\texttrademark), Cisco (IronPort\texttrademark, Email Security Appliance---ESA\texttrademark), etc., devote(d) massive R\&D investments to develop machine learning methods to automatically filter out spam in the platforms they operate (Google, Microsoft, etc.) or protect (Cisco); the companies are thus often incentivized to use patented and close-sourced solutions to maintain their competitive advantage; \textit{(ii)} related to the former point, fighting email spam is a continuous arms-race: revealing one's spam filtering technology gives out information that can be exploited by the spammers to create more sophisticated campaigns that can effectively and systematically escape detection, thus calling for more secrecy. Finally, \textit{(iii)} the accuracy of email spam detection systems deployed by these large service providers has been approaching nearly-perfect detection: a diminishing return mechanism comes into play where additional efforts to further refine detection algorithms may not warrant the costs of developing increasingly more sophisticated techniques fueling complex spam detection systems; this makes established approaches even more valuable and trusted, thus motivating the secrecy of their functioning.


\mybox{\textbf{Detecting Spam Emails}}{blue!30}{blue!10}{Email spam detection is an arms-race between attackers (spammers) and defenders (service providers). Two decades of research in the data mining and machine learning communities produced troves of techniques to tackle this problem. Some milestones include:

\paragraph{SMTP solutions} 
SMTP is the protocol at the foundation of the email exchange infrastructure. \rev{Blacklists were introduced to keep track of spam propagators \cite{caruana2012survey}. Mail servers can consult blacklisting services to determine whether to route emails to their destination. A softer version of blacklisting is \textit{greylisting}. Greylists keep track of triplets of IP addresses \textit{(sender, receiver, STMP host)} involved into an email exchange. The first time a triplet involving a dubious SMTP host appears, the exchange is denied, but the triplet is stored to authorize future exchanges. This is based on the rationale that spammers rarely retry sending spam through the same relay, and was proven effective in reducing early spam circulation \cite{caruana2012survey}.}
Another approach is keyword-based filtering: whenever the subject or the body of an email contains flagged terms (belonging to a keyword list), the SMTP service provider would not route it to its intended recipient, and flag the sending offender --- multiple offenses would lead to permanent bans. \rev{Other strategies like \textit{DomainKeys Identified Mail} (DKIM) and digital signatures are authentication methods designed to detect email spoofing and assess email provenance.}

\paragraph{Supervised learning} 
In their seminal work, \citet{drucker1999support} proposed one of the first machine learning systems for spam detection, based on Support Vector Machines (then the state of the art in terms of supervised learning). The success of supervised learning over traditional keyword-based filters demonstrated by \citet{drucker1999support} motivated the first wave of machine learning research in email spam detection. Shortly after, \citet{androutsopoulos2000experimental} showed the power of naive Bayesian anti-spam filtering:  Bayesian systems yielded state-of-the-art spam detection performance for many years. 
The advent of more sophisticated learning models, like boosting trees, set the accuracy bar higher but paradigm shifts lagged for nearly a decade. 

\paragraph{Hybrid neural systems} More recently, \citet{wu2009behavior} proposed behavior-based spam detection using combinations of simple association rules and neural networks. Given their ability to naturally handle visual information, neural network methods to detect spam were extended to multimedia content. For example, \citet{wu2005using} and \citet{fumera2006spam} proposed methods \textit{exploiting visual cues} to detect spam content injected in images embedded into emails.

\paragraph{Dedicated hardware} Networking companies are developing anti-spam appliances. Dedicated hardware can detect various types of spam, including phishing, malware, and ransomware, guaranteeing high efficiency and accuracy. For example, Cisco advertises that their Email Security Appliance (ESA\texttrademark) detects over 99.9\% of incoming spam email with lower than 1 in a million false positive rate.

}

\section{Web 2.0 or Spam 2.0? }
The new millennium brought us the Social Web, or Web 2.0, a paradigm shift with an emphasis on user-generated content and on the participatory, interactive nature of the Web experience \cite{hendler2008web}. From knowledge production (Wikipedia) to personalized news (social media) and social groups (online social networks), from blogs to image and video sharing sites, from collaborative tagging to social e-commerce, this wealth of new opportunities brought us as many new forms of spam, commonly referred to as \textit{social spam}.

Differently from spam emails, where spam can only be conveyed in one form (i.e., emails), social spam can appear in multiple forms and modi operandi. Social spam can be  in the form of textual content (e.g., a secretly-sponsored post on social media), or  multimedia (e.g., a manufactured photo on 4chan); social spam can aim at pointing users to unreliable resources, e.g., URLs to unverified information or false news Websites \cite{vosoughi2018spread}; social spam can aim at altering the popularity of digital entities, e.g., by manipulating user votes (upvotes on Reddit\texttrademark~posts, retweets on Twitter\texttrademark), and even that of physical products, e.g., by posting fake online reviews (e.g., about a product on an e-commerce Website).

\subsection{Spammy Opinions}
In the early 2000s (cf. Figure 1), the growing popularity of e-commerce Websites like Amazon and Alibaba motivated the emergence of opinion spam (a.k.a. review spam) \cite{jindal2008opinion,liu2012sentiment}. 

\rev{According to \citet{liu2012sentiment}, there are three types of spam reviews: 
(i) \textit{fake reviews}, (ii) \textit{reviews about brands only}, and (iii) \textit{non-reviews}.
The first type of spam, \textit{fake reviews},} consists of \textit{posting untruthful, or deceptive reviews on online e-commerce platforms, in an attempt to manipulate the public perception (in a positive or negative manner) of specific products or services presented on the affected platform(s)}.
Fake positive reviews can be used to enhance the popularity and positive perception of the product(s) or service(s) the spammer intends to promote, while fake negative reviews can contribute to smear the spammer's competitor(s) and their products/services. 
\rev{Opinion spam of the second type, \textit{reviews about brands only}, pertains comments on the manufacturer/brand of a product but not on the product itself---albeit genuine, according to \citet{liu2012sentiment} they are considered spam because \textit{they are not targeted at specific products and are often biased}. Finally, spam reviews of the third type, \textit{non-reviews}, are technically not opinion spam as they do not provide any opinion, they only contain generic, unrelated content (e.g., advertisement, or questions, rather than reviews, about a product). Fake reviews are, by far and large, the most common type of opinion spam, and the one that has received more attention in the research community \cite{liu2012sentiment}. Furthermore, \citet{jindal2008opinion} showed that spam of the second and third type is simple to detect and address.}

Unsurprisingly, the practice of opinion spam, \rev{and in particular fake reviews,} is widely considered as unfair and deceptive, and as such it has been subject of extensive legal scrutiny and court battles. If left unchecked, opinion spam can poison a platform and negatively affect both customers and platform providers (including incurring in financial losses for both parties, as customers may be tricked into purchasing undesirable items and grow frustrated against the platform), at the sole advantage of the spammer (or the entity they represent)---as such, depending on the country's laws, opinion spam may qualify as a form of digital fraud. 

Detecting fake reviews is complex for a variety of reasons: for example, spam reviews can be posted by fake or real user accounts. \rev{Furthermore, they can be posted by individual users or even group of users \cite{liu2012sentiment,mukherjee2012spotting}.} Spammers can deliberately use fake accounts on e-commerce platforms, created only with the scope of posting fake reviews. Fortunately, fake accounts on e-commerce platforms are generally easy to detect, as they engage in intense reviewing activity without any product purchases. An alternative and more complex scenario occurs when fake reviews are posted by real users. This tends to occur under two very different circumstances: \textit{(i)} compromised accounts (i.e., accounts originally owned by legitimate users that have been hacked and sold to spammers) are frequently re-purposed and utilized in opinion spam campaigns \cite{crawford2015survey}; \textit{(ii)} fake review markets became very popular where real users collude in exchange for direct payments to write untruthful reviews (e.g., without actually purchasing or trying a given product or service). To complicate this matter, researchers showed that fake personas, e.g., Facebook profiles, can be created and associated with such spam accounts \cite{gao2010detecting}. During the late 2000s, many online fake-review markets emerged, whose legality was battled in court by e-commerce giants. Action on both legal and technical fronts has helped mitigating the problem of opinion spam.

From a technical standpoint, a variety of techniques have been proposed to detect review spam. 
\rev{\citet{liu2012sentiment} identified three main approaches, namely supervised, unsupervised, and group spam detection. 
In \textit{supervised spam detection}, the problem of separating fake from genuine (non-fake) reviews is formulated as a classification problem. \citet{jindal2008opinion} pointed out that the main challenge of this task is to work around the shortage of labeled training data. To address this problem, the authors exploited the fact that spammers, to minimize their work, often produce (near-)duplicate reviews, that can be used as examples of fake reviews. Feature engineering and analysis was key to build
informative features of genuine and fake reviews, enriched by features of the reviewing users and the reviewed products. Models based on Logistic Regression have been proven successful in detecting untruthful opinions in large corpora of Amazon reviews \cite{jindal2008opinion}.} Detection algorithms based on Support Vector Machines or Naive Bayes models generally perform well (above 98\% accuracy) and scale to production systems \cite{mukherjee2013spotting}. These pipelines are often enhanced by human-in-the-loop strategies, where annotators recruited through Amazon Mechanical Turk (or similar crowd-sourcing services) manually label subsets of reviews to separate genuine from fake ones, to feed online learning algorithms so to constantly adapt to new strategies and spam techniques \cite{liu2012sentiment,crawford2015survey}. 

\rev{\textit{Unsupervised spam detection} was used both to detect spammers as well as for detecting fake reviews. \citet{liu2012sentiment} reported on methods based on detecting anomalous behavioral patterns typical of spammers.  Models of spam behaviors include \textit{targeting products}, \textit{targeting groups (of products or brands)}, \textit{general and early rating deviations} \cite{liu2012sentiment}. Methods based on association rules can capture atypical behaviors of reviewers, detecting anomalies in reviewers' confidence, divergence from average product scores, entropy (diversity or homogeneity) of attributed scores, temporal dynamics, etc. \cite{xie2012review}.
For what concerns the unsupervised detection of fake reviews, } 
linguistic analysis was proved useful to identify stylistic features of fake reviews e.g., language markers that are over- or under-represented in fake reviews. Opinion spam to promote products, for example, exhibits on average three times fewer mentions of social words, negative sentiment, and long words ($>6$ letters) than genuine reviews, while containing twice more positive terms and references to self than formal texts \cite{crawford2015survey}. 

\rev{
Concluding, \textit{group spam detection} aims at identifying signatures of collusion among spammers
\cite{mukherjee2012spotting}. Collective behaviors such as spammers' coordination can emerge by using combinations of \textit{frequent pattern mining} and \textit{group anomaly ranking}. In the first stage, the algorithm proposed by \citet{mukherjee2012spotting} identifies groups of reviewers who all have reviewed a same set of products---such groups are flagged as potentially suspicious. Then, anomaly scores for individual and group behaviors are computed and aggregated, accounting for indicators that measure the group burstiness (i.e., writing reviews in short timespan), group reviews similarity, etc. Groups are finally ranked in terms of their anomaly scores \cite{mukherjee2012spotting}. 
}



\subsection{The Rise of Spam Bots}
Up to the early 2000s,  most of the spam activity was still coordinated and carried out, at least in significant part, by human operators: email spam campaigns, Web link farms, fake reviews, etc. all rely on human intervention and coordination. In other words, these spam operations scale at a (possibly significant) cost. 
\rev{With the rise in popularity of online social network and social media platforms  (see Figure 1), new forms of spam started to emerge at scale. One such example is social link farms \cite{ghosh2012understanding}: similarly to Web link farms, whose goal is to manipulate the perception of popularity of a certain Website by artificially creating many  pointers (hyperlinks) to it, in social link farming spammers create online personas with many artificial followers. This type of spam operation requires creating thousands (or more) of accounts that will be used to follow a target user in order to boost its apparent influence. Such ``disposable accounts'' are often referred to as \textit{fake followers} as their purpose is solely to participate in such link-farming networks. In some platforms, link farming was so pervasive that spammers reportedly controlled millions of fake accounts \cite{ghosh2012understanding}.  Link farming introduced a first level of automation in social media spam, namely the tools to automatically create large swaths of social media accounts.}

In the late 2000s, social spam obtained a new potent tool to exploit: bots (short for software robots, a.k.a. social bots). In my 2016 CACM review titled \textit{The Rise of Social Bots} \cite{ferrara2016rise}, I noted that ``bots have been around since the early days of computers'': examples of bots include \textit{chatbots}, algorithms designed to hold a conversation with a human, \textit{Web bots}, to automate the crawling and indexing of the Web, \textit{trading bots}, to automate stock market transactions, and much more. Although isolated examples exist of such bots being used for nefarious purposes, I am unaware of any reports of systematic abuse carried out by bots in those contexts.

A social bot is a new breed of ``computer algorithm that automatically produces content and interacts with humans on the Social Web, trying to emulate and possibly alter their behavior.''
Since bots can be programmed to carry out arbitrary operations that would otherwise be tedious or time-consuming (thus expensive) for humans, they allowed to scale spam operations on the Social Web to an unprecedented level. 
Bots, in other words, are the dream spammers have been dreaming of since the early days of the Internet: they allow for personalized, scalable interactions, increasing the cost effectiveness, reach, and plausibility of social spam campaigns, with the added advantage of increased credibility and the ability to escape detection achieved by their human-like disguise.
Furthermore, with the democratization and popularization of machine learning and AI technologies, the entry barrier to creating social bots has significantly lowered \cite{ferrara2016rise}. 
Since social bots have been used in a variety of nefarious scenarios (see \textit{Sidebar: Social Spam Applications}), from the manipulation of political discussion, to the spread of conspiracy theories and false news, and even by extremist groups for propaganda and recruitment, the stakes are high in the quest to characterize bot behavior and detect them \cite{ICWSM1715587}.\footnote{\rev{It should be noted that bots are not used exclusively for nefarious purposes: for example, some researchers used bots for positive health behavioral interventions \cite{ferrara2016rise}. Furthermore, it has been noted that the most problematic aspect of nefarious bots is their attempt to deceive and disguise themselves as human users \cite{ferrara2016rise}: however, many bots are labeled as such and may provide useful services, like live-news updates, etc.}}

Maybe due to their fascinating morphing and disguising nature, spam bots have  attracted the attention of the AI and machine learning research communities: the arms-race between spammers and detection systems yielded technical progress on both the attacker's and the defender's technological fronts. Recent advancements in Artificial Intelligence (especially Artificial Neural Networks) fuel bots that can generate human-like natural language  and interact with human users in near real time \cite{ferrara2016rise, ICWSM1715587}. On the other hand, the cyber-security and machine-learning communities came together to develop techniques to detect the signature of artificial activity of bots and social network sybils \cite{yang2014uncovering, ferrara2016rise}.

In \cite{ferrara2016rise}, we flashed out techniques used to both create spam bots, and detect them. Although the degree of sophistication of such bots, and therefore their functionalities, varies vastly across platforms and application domains, commonalities also emerge. Simple bots can do unsophisticated operations, such as posting content according to a schedule, or interact with others according to pre-determined scripts, whereas complex bots can motivate their reasoning and react to further human scrutiny. Beyond anecdotal evidence, there is no systematic way to survey the state of AI-fueled spam bots and consequently their capabilities---researchers adjust their expectations based on advancements made public in AI technologies (with the assumptions that these will be abused by spammers with the right incentives and technical means), and based on proof-of-concept tools that are often originally created with other non-nefarious purposes in mind (one such example is the so-called \textit{DeepFakes}, discussed more later). 

In the \textit{Sidebar: Social Spam Applications}, I highlight some of the domains where bots made the headlines: one such example is the wake to the 2016 U.S. presidential election, during which Twitter and Facebook bots have been used to sow chaos and further polarize the political discussion \cite{bessi2016social}. Although it is not always possible for the research community to pinpoint the culprits, the research of my group, among many others, contributed to unveil anomalous communication dynamics that attracted further scrutiny by law enforcement and were ultimately connected to state-sponsored  operations (if you wish, a form of social spam aimed at influencing individual behavior). Spam bots operate in other highly-controversial conversation domains: in the context of public health, they promote products or spread scientifically unsupported claims \cite{ferrara2015manipulation, allem2017cigarette}; they have been used to create spam campaigns to manipulate the stock market \cite{ferrara2015manipulation}; 
finally, bots have also been used to penetrate online social circles to leak personal user information \cite{gao2010detecting}.


\mybox{\textbf{Social Spam Applications}}{blue!30}{blue!10}{
\paragraph{Political manipulation} In a peer-reviewed study published on November 7, 2016 \cite{bessi2016social} (the day before the U.S. presidential election), I unveiled a massive-scale spam operation affecting the American political Twitter. With the aid of \textit{Botometer}, an AI system that leverages over a thousand features to separate bots from humans \cite{ICWSM1715587}, hundreds of thousands of bots were identified. By studying the activity signatures of these bots, I noted that they were being retweeted at the same rate than human users, which may have contributed to the spread of political misinformation \cite{vosoughi2018spread}. Since most of these bots aimed at sowing chaos, their presence may have inflamed and further polarized the political conversation, with unknown consequences on the integrity of the vote. Since then, dozens of studies corroborated these results; many other studies, before and after mine, showed the perils associated with social spam campaigns in political domains. \rev{Most recently, the emerging phenomenon of \textit{fake news spreading} attracted a lot of attention. \citet{vosoughi2018spread} investigated the role of social media, as well as bots, in the spread of true and false news: the authors showed that humans are more likely to share false stories inspired by fear, disgust, and surprise. This suggests that conditioning and manipulation operations online can affect human behavior.}

\paragraph{Public heath}
Conspiracy and denialism are endemic of social networks. Spam in public health discussions has become commonplace for social media: in a recent study, for example, my team highlighted how bots are used to promote electronic cigarettes as cessation devices with health benefits, a fact not definitively corroborated by science  \cite{allem2017cigarette}.
The use of bots to carry out anti vaccination campaigns has been the subject of investigation of a DARPA Challenge in 2016 \citep{subrahmanian2016darpa}.

\paragraph{Stock market} Automatic trading algorithms leverage information from social media to predict stock prices. Using bots, spam campaigns have been carried out to give the false impression that certain stocks were spoken positively about on Twitter, successfully tricking trading algorithms into buying them in a pump-and-dump scheme unveiled by the U.S. Securities and Exchange Commission
 (SEC) in 2015 \cite{ferrara2015manipulation}.

\paragraph{Data leaks}
Social platforms enable  the often unwilling disclosure of private user information. 
\rev{A recent study showed that over a third of content shared on Facebook has the default public-visibility privacy settings \cite{liu2011analyzing}. The amount of content accessible to undesirable users may be}
 even higher when considering privacy settings that allow one's friends to access  private information and preferences:  Research showed that most users indiscriminately accept friendship connections on Facebook \cite{gao2010detecting}. 
Spam bots can inject themselves into tightly-connected communities, by leveraging the weak-tie structure of online social networks \cite{demeo2014facebook}, and obtain private user information on large swaths of users. 
\rev{Phishing is also responsible for data leaks. Attacks based on short-URLs are popular on social media: they can hide the true identity of the spammers and have been proven  effective to steal personal data \cite{chhabra2011phi,ghosh2012understanding}.}
}

\section{AI SPAM}
Artificial Intelligence has been advancing at vertiginous speed, revolutionizing many fields including spam. 
Beyond powering conversational agents such as social bots, as discussed above, AI systems can be used, beyond their original scope, to fuel spam operations of different sorts. I will refer to this phenomenon next as \textit{spamming with AI}, hinting to the fact that AI is used as a tool to create new forms of spams. 
However, given their sophistication, AI systems can themselves be subject of spam attacks. I will refer to this new concept as \textit{spamming into AI}, suggesting that AIs  can be manipulated, and even compromised, by spammers (or attackers in a broader sense) to exhibit anomalous and undesirable behaviors.

\subsection{Spamming with AI}

\begin{figure}[t]
\includegraphics[width=\columnwidth]{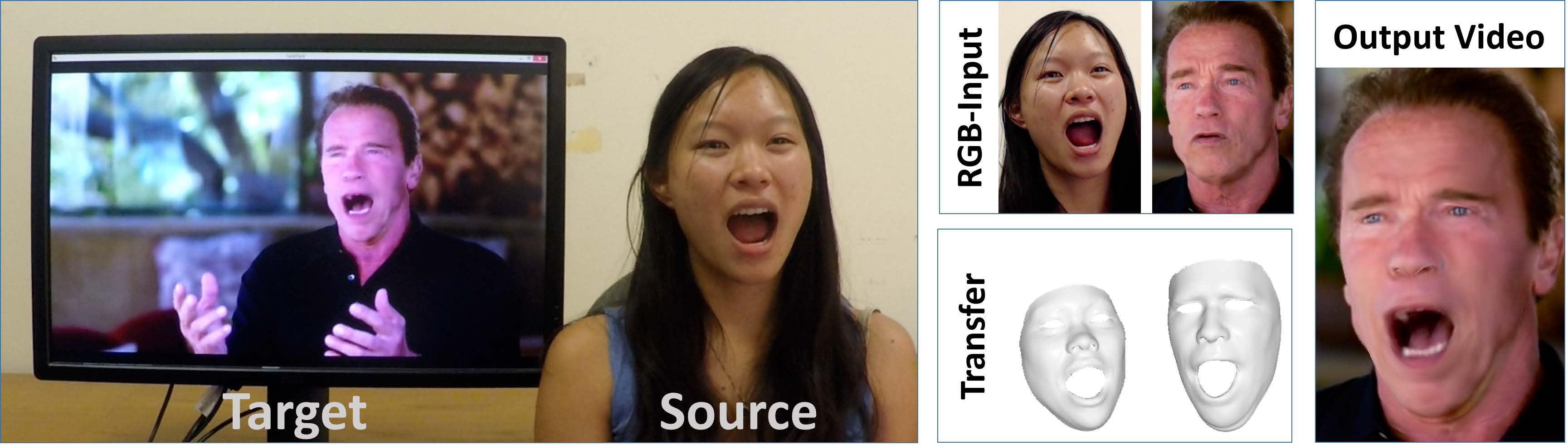}
\caption{Video sequence real-time reenactment using AI (From \citep{thies2016face}). This proof-of-concept technology could be abused to create AI-fueled multimedia spam.}\label{fig:teaser}
\end{figure}

Advancements in computer vision, augmented and virtual realities are projecting us in an era where the boundary between reality and fiction is increasingly more blurry. 
Proofs-of-concept of AIs capable to analyze and manipulate video footages, learning patterns of expressions, already exist: \citeauthor{suwajanakorn2017synthesizing} \cite{suwajanakorn2017synthesizing} designed a deep neural network to map any audio into mouth shapes and convincing facial expressions, to impose an arbitrary speech on a video clip of a speaking actor, with results hard to distinguish, to the human eye, from genuine footage. \citeauthor{thies2016face}~\cite{thies2016face} showcased a technique for real-time facial reenactment, to convincingly re-render the synthesized target face on top of the corresponding original video stream (see Figure \ref{fig:teaser}). These techniques, and their evolutions \cite{kim2018deep}, have been then exploited to create so-called \textit{DeepFakes}, face-swaps of celebrities into adult content videos that surfaced on the Internet by the end of 2017. 
Such techniques have also already been applied to the political domain, creating fictitious video footage re-enacting   Obama,\footnote{See \url{https://grail.cs.washington.edu/projects/AudioToObama}} Trump,  and Putin,\footnote{See \url{http://niessnerlab.org/projects/thies2016face.html}} among several world  leaders \cite{kim2018deep}.
Concerns about the ethical and legal conundra of these new technologies have been already expressed \cite{chesney2018deep}.

In the future, these technologies may be abused by well-resourced spammers to create AIs pretending to be human.
Another example: Google recently demonstrated the ability to deploy an
AI (Google Duplex\texttrademark) in the real world to act as a virtual assistant, seamlessly interacting with human interlocutors over the phone:\footnote{\url{https://ai.googleblog.com/2018/05/duplex-ai-system-for-natural-conversation.html}} such technology may likely be re-purposed to carry out massive scale spam-call campaigns. Other forms of future spam with AI may use augmented or virtual reality agents, so-called Digital Humans, to interact with humans in digital and virtual spaces, to promote products/services, and in worse-case scenarios to carry out nefarious campaigns similar to those of today's  bots, to manipulate and influence users.

\subsection{Spamming into AI}
AIs based on Artificial Neural Networks (ANNs) are sophisticated systems whose functioning can sometimes be too complex to explain or debug. For such a reason, ANNs can be easy preys of various forms of attacks, including spam, to elicit undesirable, even harmful system's behaviors. 
An example of spamming into AI can be \textit{bias exacerbation}: one of the major problems of modern-days AIs (and, in general, of supervised learning approaches based on Big Data) is that biases learned from training data will propagate into predictions. The problem of bias \cite{baeza2018bias}, especially in AI, is under the spotlight and is being tackled by the computing research community.\footnote{\url{https://www.technologyreview.com/s/608986/forget-killer-robotsbias-is-the-real-ai-danger}}
One way an AI can be maliciously led to learn biased models is deliberately injecting spam---here intended as unwanted information---into the training data: this may lead the system to learn undesirable patterns and biases, which will affect the AI system's behavior in line with the intentions of the spammers.

\begin{figure}[t]
\includegraphics[width=\columnwidth]{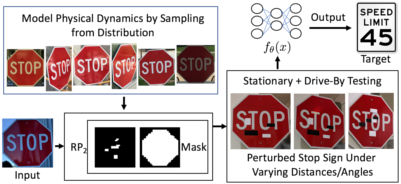}
\caption{Physical-world attacks onto AI visual classifier (From \cite{eykholt2018robust}). Similar techniques could be abused to inject unwanted spam into AI and trigger anomalous behaviors.}\label{fig:real}
\end{figure}

An alternative way of spamming into AI is the manipulation of test data. If an attacker has a good understanding of the limits of an AI system, for example by having access to its training data and thus the ability to learn strength and weakness of the learned models, attacks can be designed to lure the AI into an undesirable state.
Figure \ref{fig:real} shows an example of a physical-world attack that affects an AI system's behaviors in anomalous and undesirable ways \cite{eykholt2018robust}: in this case, a deep neural network for image classification (which may have been used, for example, to control an autonomous vehicle) is tricked by a ``perturbed'' stop sign mistakenly interpreted as a speed limit sign---according to the expectation of the attacker. 
Spam test data may be displayed to a victim AI system to lure it into behaving according to a scripted plot based on weaknesses of the models and/or of its underlying data. 
The potential applications of such type of spam attacks can be in medical domains (e.g., deliberate misreading of scans), autonomous mobility (e.g., attacks on the transportation infrastructure or the vehicles), and more.
Depending on the pervasiveness of AI-fueled systems in the future, the questions related to spamming into AI may require the immediate attention of the research community. 

\section{Recommendations}
Four decades have passed since the first case of email spam was reported by 400 ARPANET users (cf. Figure 1). 
While some prominent computer scientists (including Bill Gates) thought that spam would  quickly be solved and soon remembered as a problem of the past \cite{cranor1998spam}, we have witnessed its evolution in a variety of forms and environments. Spam feeds itself of (economic, political, ideological, etc.) incentives and of new technologies, both of which there is no shortage of, and therefore it is likely to plague our society and our systems for the foreseeable future.

It is therefore the duty of the computing community to enact policies and research programs to keep fighting against the proliferation of current and new forms of spam. I conclude suggesting three maxims that may guide future efforts in this endeavor: 

\paragraph{1. Design technology with abuse in mind}
Evidence seems to suggest that, in the computing world, new powerful technologies are oftentimes abused beyond their original scope. Most modern-days technologies, like the Internet, the Web, email, and social media, have not been designed with built-in protection against attacks or spam. However, we cannot perpetuate a naive view of the world that ignores ill-intentioned attackers: new systems and technologies shall be designed from their inception with abuse in mind. 

\paragraph{2. Don't forget the arms-race}
The fight against spam is a constant arms-race between attackers and defenders, and as in most adversarial settings, the party with the highest stakes will prevail: since with each new technology comes abuse, researchers shall anticipate the need for counter-measures to avoid being caught unprepared when spammers will abuse their newly-designed technologies. 

\paragraph{3. Blockchain technologies} 
The ability to carry out massive spam attacks in most systems exists predominantly due to the lack of authentication measures that reliably guarantee the identity of entities and the legitimacy of transactions on the system. The Blockchain as a proof-of-work mechanism to authenticate digital personas (including in virtual realities), AIs, etc. may prevent several forms of spam and mitigate the scale and impact of others.\footnote{\rev{It is worth noting that proof-of-work has been proposed to prevent spam email in the past, however its feasibility remains debated, especially in its original non Blockchain-based implementation \cite{laurie2004proof}.}} 

\bigskip
Spam is here to stay: let's fight it together!
\small
\begin{acks} 
The author would like to thank current and former members of the USC Information Sciences Institute's MINDS research group, as well as of the Indiana University's CNetS group, for invaluable research collaborations and discussions on the topics of this work.
The author is grateful to his research sponsors including the \grantsponsor{AFOSR}{Air Force Office of Scientific Research}{} (AFOSR), award  \grantnum{FakeNews}{FA9550-17-1-0327}, and the \grantsponsor{DARPA}{Defense Advanced Research Projects Agency}{} (DARPA), contract  \grantnum{SocialSim}{W911NF-17-C-0094}. 
\end{acks}

\bibliographystyle{ACM-Reference-Format}
\bibliography{ferrara-bibliography}

\begin{figure*}
\includegraphics[width=2.1\columnwidth]{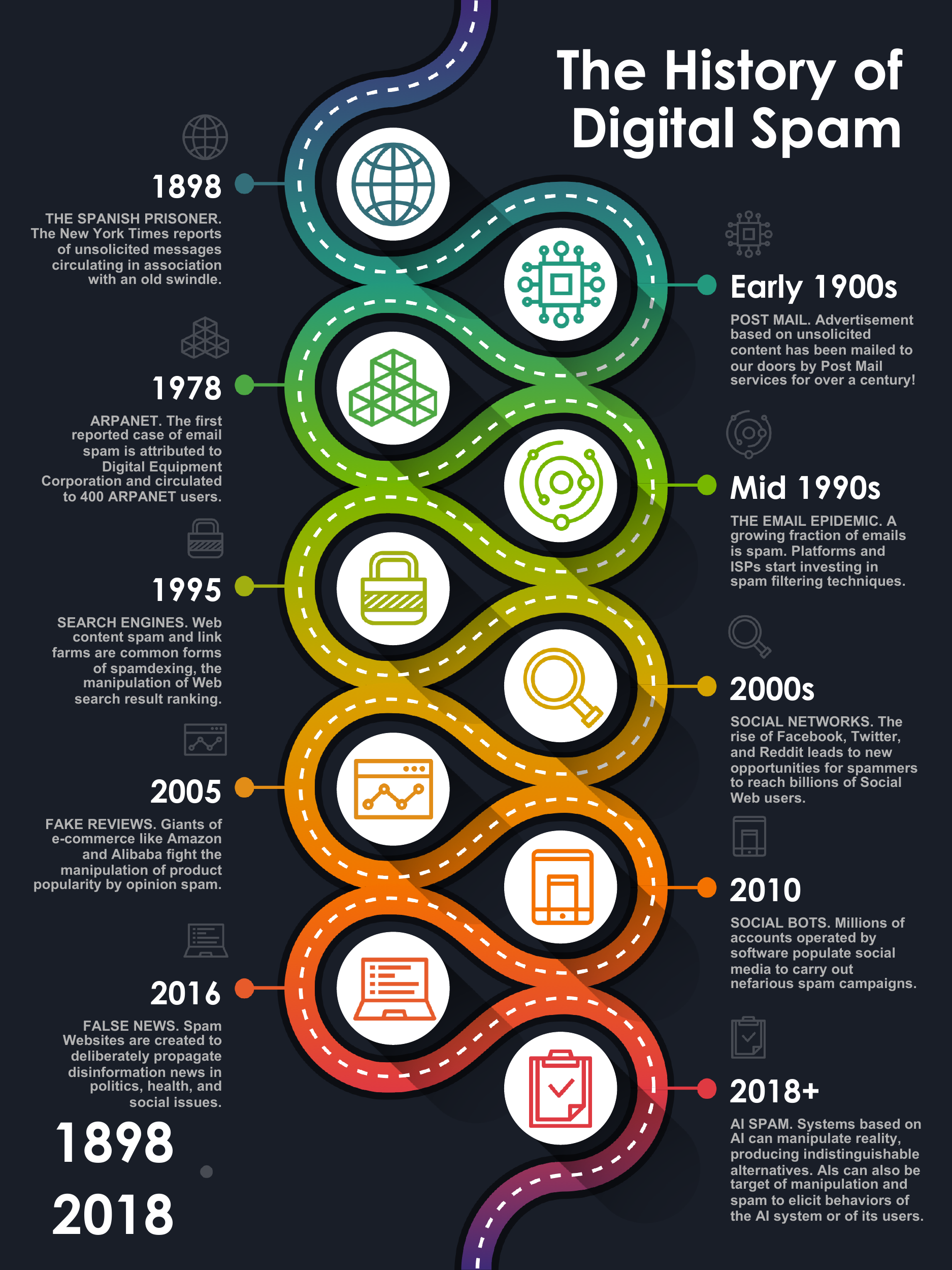}
\label{fig:cover}\vspace*{-5mm}
\end{figure*}

\end{document}